\documentstyle[twoside,fleqn,npb]{article}

\newcommand{\ep}{\varepsilon}

\title{Where do the tedious products of $\zeta$'s come from?}

\author{D. J. Broadhurst
\address{Department of Physics and Astronomy,\\
The Open University, Milton Keynes MK7 6AA, U.K.}}

\begin{document}

\begin{abstract}
Lamentably, the full analytical content of the $\ep$--expansion
of the master two--loop two--point function, with arbitrary self--energy
insertions in $4-2\ep$ dimensions, is still unknown. Here we show that
multiple zeta values (MZVs) of weights up to 12 suffice through $O(\ep^9)$.
Products of primitive MZVs are generated by a processes of
``pseudo--exponentiation'' whose combinatorics faithfully accord
with expectations based on Kreimer's modified shuffle product
and on the Drinfeld--Deligne conjecture. The existence
of such a mechanism, relating thousands of complicated rational numbers,
enables us to identify precise and simple combinations of MZVs
specific to quantum field theories in even numbers of spacetime dimensions.
\end{abstract}

\maketitle

\section{Master two--loop two--point function}

The object of study in this contribution is the two--loop integral
\begin{eqnarray}
I&=&\frac{p^{2(D/2-\alpha_6)}}{\pi^{D}}
\int\int{{\rm d}^{D}k{\rm d}^{D}l\over P}\label{I}\\
P&=&(k-p)^{2\alpha_1}
(l-p)^{2\alpha_2}
(k-l)^{2\alpha_3}
l^{2\alpha_4}
k^{2\alpha_5}
\label{P}
\end{eqnarray}
with
$\alpha_6\equiv3D/2-\alpha_1-\alpha_2-\alpha_3-\alpha_4-\alpha_5$
determining the dependence
on the number, $D=4-2\ep$, of spacetime dimensions.
It is clearly independent
of the norm, $p^2$, of the external momentum $p$. Remarkably, it
has a 1440--element symmetry group~\cite{1440}, corresponding
to all permutations of 6 linear combinations~\cite{Z6} of the
indices $\alpha_j$,
combined with the total reflection $\alpha_j\to D/2-\alpha_j$.

With parameters $\alpha_{j}=1+n_j\ep$, resulting from insertions
on internal lines, the Taylor expansion of $I=6\zeta(3)+O(\ep)$
is known~\cite{BGK} to involve multiple zeta values.

The MZV $\zeta(5,3)\equiv\sum_{m>n>0}1/m^5n^3$,
with weight 8 and depth 2, appears in $I$ at order $\ep^5$ and is
also present in the six--loop beta function of $\phi^4$ theory~\cite{BK15}.

Similarly, $\zeta(3,5,3)\equiv\sum_{l>m>n>0}1/l^3m^5n^3$,
with weight 11 and depth 3, appears at order $\ep^8$
and in $\phi^4$ theory at 7 loops.

In general, we expect $L$--loop
counterterms to include MZVs of weights $w\le2L-3$,
of which the expansion of $I$ through order $\ep^{2L-6}$ is
to be taken as a strong diagnostic.

It is one of the many scandals of our limited understanding of the
analytical content of perturbative quantum field theory (QFT)
that, despite many years of intense effort, we still do not
know whether MZVs suffice for even the Taylor expansion of the
two--loop integral~(\ref{I}).

Here we exhibit remarkable structure in the dependence
of the $\sum_{k=0}^9{5+k\choose5}=5005$ Taylor coefficients
through $O(\ep^9)$ on the
\begin{equation}
1+1+2+2+3+4+5+7+9+12=46
\label{46}
\end{equation}
independent MZV structures
from weight 3 to 12. In particular, MZV
structures that are imprimitive products
will be generated by a process of ``pseudo--exponentiation''
whose very existence appears to me to be little short of
miraculous and whose details
resonate strongly both with Dirk Kreimer's
combinatorically modified shuffle product,
discussed in these proceedings, and also with the
Drinfeld--Deligne conjecture, namely that the
Grothendieck--Teichm\"uller algebra has precisely
one generator of each odd degree greater than unity,
and none of even degree.

\section{Choices of primitive MZVs}

As Don Zagier observed, the Drinfeld--Deligne conjecture leads
to an enumeration $P(n)=P(n-2)+P(n-3)$ of
{\bf Q}--linearly (i.e.\ rationally)
independent MZV structures of weight $n$,
seeded by $P(1)=0$, $P(2)=P(3)=1$. This generates
the Padovan numbers $P(3)$ to $P(12)$ in~(\ref{46}).
The filtration of products into primitives then requires
\begin{equation}
M(w)=\frac{1}{w}\sum_{n|w}\mu(w/n)Q(n)
\end{equation}
primitives to be chosen at weight $w$, where
$Q(n)=Q(n-2)+Q(n-3)$ is the Perrin sequence,
seeded by $Q(1)=0$, $Q(2)=2$, $Q(3)=3$, and 
the M\"obius function $\mu$
ensures that
\begin{equation}
\prod_{n>0}(1-x^n)^{-M(n)}={1\over1-x^2-x^3}.\label{filt}
\end{equation}
At weight $w=2$ we have the primitive $\pi^2$ (which
mathematicians sometimes relegate to $w=0$) and
at each odd weight $2n+1>1$ we have the primitive $\zeta(2n+1)$.
That leaves us with one more primitive MZV to choose at each
of the weights 8, 10 and 11. We might take these as
$\zeta(6,2)$, $\zeta(8,2)$ and $\zeta(8,2,1)$. At weight
$w=12$ we need a primitive of depth 2, such as $\zeta(10,2)$,
and one of depth 4, such as $\zeta(4,4,2,2)$.

{}From a mathematical point of view, these are entirely
adequate choices of primitives, since every MZV
of weight $w\le12$ is expressible as a rational
linear combination of products of the filtration.

However, we are concerned here with QFT, where more
specific combinations of MZVs are found to
occur in counterterms~\cite{BK15}. I shall
sketch how I was led, by Kreimer's fertile ideas~\cite{K,CK,BK},
and by the Drinfeld--Deligne conjecture, to the following
set of QFT primitives up to weight 12.
\begin{enumerate}
\item At weights $w=2,3,5,7,9,11$, take $\zeta(w)$.
\item At $w=8$, take $\zeta(5,3)$.
\item At $w=10$, take $\zeta(7,3)+3\zeta^2(5)$.
\item At $w=11$, take $\zeta(3,5,3)-\zeta(3)\zeta(5,3)$.
\item At $w=12$, take $\zeta(7,5)$ and the combination
of {\em alternating\/} Euler sums below.
\end{enumerate}
The choice of a second primitive at weight 12 is
\begin{equation}
Z_2(12)=\zeta(\overline7,5)-
\zeta(\overline5,7)+
\zeta(\overline5)\zeta(\overline7)
\label{other}
\end{equation}
comprising Euler sums of the form
\begin{eqnarray}
\zeta(\overline{a},b)&=&\sum_{m>n>0}{(-1)^m\over m^a}{1\over n^b}
\label{z2a}\\
\zeta(\overline{a})&=&\sum_{n>0}{(-1)^n\over n^a}
=(2^{1-a}-1)\zeta(a)
\label{za}
\end{eqnarray}
with alternation of sign notated by a bar above the corresponding index.

With these very specific choices, products in the Taylor
expansion of~(\ref{I}) through order $\ep^9$
will be generated by pseudo--exponentiation. The choice of
primitives which makes this possible is essentially unique.
The existence of such a choice is truly remarkable.
For example, the choice of primitive in~(\ref{other})
was determined by a tiny subset of the available data,
whereupon transformation to the appropriate basis reduced a
file of many megabytes of tedious products of $\zeta$'s
to a single line of code.

\section{The master function}
The key to the expansion of~(\ref{I}) is the function~\cite{BGK}
\begin{eqnarray}
S(a,b,c,d)&=&\frac{\pi\cot\pi c}{H(a,b,c,d)}
-\frac{1}{c}\nonumber\\
&-&\frac{b+c}{b c}F(a+c,-b,-c,b+d)\label{S}
\end{eqnarray}
with a hypergeometric series of the form
\begin{equation}
F(a,b,c,d)=
\sum_{n=1}^\infty\frac{(-a)_n(-b)_n}
{(1+c)_n(1+d)_n},\label{F}
\end{equation}
where $(a)_n\equiv\Gamma(a+n)/\Gamma(a)$, and with
a ratio of products of Gamma functions of the form
\begin{equation}
H(a,b,c,d)={G(\{a,b,c,d,a+b+c+d\})
\over G(\{a+c,a+d,b+c,b+d\})}\label{H}
\end{equation}
where $G({\cal S})\equiv\prod_{a\in{\cal S}}\Gamma(1+a)$.

If the propagators in two adjacent internal lines of
the master two--loop diagram have unit exponents
(i.e.\ no dressing by self--energy insertions)
then the recurrence relations from integration by parts
may be solved in terms of two instances~\cite{BGK}
of the construct~(\ref{S}), which solves
$$
aS(a,b,c,d)=1
+{(a+c)(a+d)\over a+b+c+d}S(a-1,b,c,d)
$$
in the presence of the symmetries
\begin{equation}
S(a,b,c,d)=S(b,a,c,d)=-S(c,d,a,b).
\label{sym}
\end{equation}

Moreover, the 1440--fold symmetry
of the master diagram enables one to reconstruct the 9th--order
Taylor expansion of $I$, with 6
arguments, from the 11th--order expansion of $S$,
with only 4 arguments. If there is any new type of number,
beyond the MZVs that are generated by the $\ep$--expansions of
Pochhammer symbols in~(\ref{F}), then it will show up only
at weights greater than 12.

\section{Pushdown at weight 12}

The technology for developing the Taylor expansion of $S$ through
weight 11 was developed in~\cite{BGK}. At weight 12 we now
encounter a remarkable phenomenon: the depth--4 MZV $\zeta(4,4,2,2)$
is primitive within the confines of the theory of MZVs,
but widening the analysis to include alternating Euler
sums, one discovers that it may be eliminated in favour of
a depth--2 sum, such as $\zeta(\overline9,\overline3)
=\sum_{m>n>0}(-1)^{m+n}/m^9n^3$. Specifically, the combination
$\zeta(4,4,2,2)-(8/3)^3\zeta(\overline9,\overline3)$
is reducible to 11 weight--12 terms in the MZV basis,
thus showing that $\zeta(\overline9,\overline3)$ can
furnish the 12th.

Here one sees another scandal of our limited understanding.
This ``pushdown'' from depth--4 MZVs to depth--2 Euler sums
was found empirically using David Bailey's implementation
of Helaman Ferguson's PSLQ algorithm, which turns high
precision numerical evaluations into immensely probable
integer relations. In this case, the effort was infinitesimal,
in comparison with the PSLQ investigations in~\cite{bb}.
Nevertheless, no--one seems to have any idea of how to prove
such pushdowns rigorously, apart from labouring to solve the
huge complex of linear relations between $3^{12}=531441$
words of 12 letters, taken from the alphabet
$A={\rm d}x/x$, $B={\rm d}x/(1-x)$, $C=-{\rm d}x/(1+x)$ for
the iterated integrals corresponding to Euler sums.

In terms of CPUtime, it takes but a second of PSLQ work
to find the simple coefficient $(8/3)^3$
of $\zeta(\overline9,\overline3)$ in the pushdown of $\zeta(4,4,2,2)$.
Yet it would appear to require a huge expense of computer algebra,
feasible perhaps only with Jos Vermaseren's {\sc form}, to obtain this
reduction by systematic application of the shuffle
algebras of nested sums and iterated integrals in the $\{A,B,C\}$
alphabet~\cite{bbb,bbbl}.

To develop the Taylor expansion of $S$ to weight 12, PSLQ was used
to find 139 such reductions of MZVs to a basis involving
$\zeta(\overline9,\overline3)$.

\section{Pseudo--exponentiation}

After achieving the pushdown to depth 2,
one is left with something that seems hugely
indigestible, namely the reduction of
$\sum_{k=1}^{11}{3+k\choose 3}=1364$
Taylor coefficients of $S$ to {\bf Q}--linear combinations of
$\sum_{k=1}^{11} P(k+1)=47$ terms formed by products of primitives.
The data comprises many thousands of
rational numbers which are typically the
ratios of 10--digit integers.
Some reduction of this complexity
is achieved by using the symmetries~(\ref{sym}). For example the
leading term in $S(a,b,c,d)$ is simply $(a+b-c-d)\zeta(2)$
and this cancels when one combines the two instances of $S$
in $I=6\zeta(3)+O(\ep)$. Even after taking account of such
obvious symmetries, several thousand product terms
remain to be explained.

Discussion with Dirk Kreimer suggested that perhaps some
of the tedious products of primitives might follow a pattern.
The basic idea was that some simple structure might be delicately
``composed'' with an expression relatively free of such products,
so that, for example, one might write $S=E\vee R$,
where $E$ is made of Gamma functions (and hence exponentiates
a series of $\zeta$'s), $R$ is far simpler than $S$,
and the rule of composition, denoted by $\vee$, has the distinctive
QFT property of counting like objects and then modifying the
ordinary process of multiplication by an appropriate symmetry factor.

There was an obvious candidate for $E$: the reciprocal of $H$
in the first term of~(\ref{S}), corresponding to
exponentiation of
$$-\log H(a,b,c,d)=\sum_{s>1}(-1)^s\zeta(s){N(s)-D(s)\over s},$$
$$N(s)=a^s+b^s+c^s+d^s+(a+b+c+d)^s,$$
$$D(s)=(a+c)^s+(a+d)^s+(b+c)^s+(b+d)^s.$$
However it soon became clear that the powers of $\pi^2$
in $\log H$ can play no role in the composition $E\vee R$.
This seemed to chime well with the Drinfeld--Deligne
conjecture. Hence the Ansatz
\begin{equation}
E(a,b,c,d)=\left(H(-a,-b,-c,-d)\over H(a,b,c,d)\right)^{1/2}
\end{equation}
was made, since it exponentiates only odd $\zeta$'s.

The challenge was then to find a product--free expansion
of the form
\begin{eqnarray}
R(a,b,c,d)&=&\sum_{w=2}^{12} \zeta(w)J_w(a,b,c,d)\nonumber\\
&+&\sum_{w=8,10,11}Z(w)K_w(a,b,c,d)\nonumber\\
&+&\sum_{k=1,2}Z_k(12)K_{12,k}(a,b,c,d)\label{R}
\end{eqnarray}
involving the 11 single sums $\zeta(w)=\sum_{n>0}1/n^w$,
with $w\in[2,12]$, a single additional primitive MZV at each of the
weights $w=8,10,11$, and two such primitives at $w=12$.
Since all 16 of these terms are present
in $S$, and $E=1+O(\ep^3)$ does not exponentiate powers
of $\pi^2$, this is the minimal product--free Ansatz for $R$.
What was far from clear is that there exists a well--defined procedure
$E\vee R$ that can generate the remaining $47-16=31$ product
structures in $S$ from the $11+5=16$ single--sum and primitive terms
in $R$. Bearing in mind that a weight--$w$ product term in $S$
entails all ${w+2\choose3}$ monomials of degree $w-1$ in 4 variables,
the odds against finding a composition procedure
appeared at first so overwhelming as to make such an enterprise
foolish.

However, following the well tried maxim that ``QFT is far smarter
than we are'', I looked to see what happens when one takes
the composition to be that of simple multiplication. This
faithfully reproduced product terms such as $\zeta(3)\zeta(2)$
and $\zeta(3)\zeta(4)$, but to Dirk Kreimer's great delight
gave the wrong result for $[\zeta(3)]^2$, whose true dependence
on the 56 monomials of 5th order in 4 variables
is proportional to $(N(3)-D(3))J_3$
yet has precisely {\em half\/} the coefficient that would
be predicted by straightforward multiplication. In other words
the composition $E\vee R$ must take account of symmetry
factors: if $[\zeta(n)]^p/p!$ from the exponential $E$
is composed with $\zeta(n)$ from the product--free expansion $R$,
the contribution to $E\vee R$ must be taken as $[\zeta(n)]^{p+1}/(p+1)!$.
Thus to generate
the 165 monomials in the $[\zeta(3)]^3$ term of $S$
we must modify the result of naive multiplication by
the factor $2!/3!=1/3$. That aspect of pseudo--exponentiation
had been eagerly anticipated.

A second aspect came as a surprise,
though in retrospect it signals that
QFT is in deep accord with the Drinfeld--Deligne conjecture.
The first example occurs
at weight 8, where we know that we must include a new
primitive, such as $\zeta(5,3)$, in $Z(8)$. To a mathematician,
such a primitive could include an arbitrary rational multiple
of $\zeta(5)\zeta(3)$. But we seek to generate all monomials
found to be multiplying $\zeta(5)\zeta(3)$ in $S$, using
the composition $E\vee R$. There are, at most, three
sources: $\zeta(5)\vee\zeta(3)$, $\zeta(3)\vee\zeta(5)$
and, possibly, a multiple of $\zeta(5)\zeta(3)$ from $1\vee Z(8)$.
The functional dependencies on $a,b,c,d$ of all three contributions
are predetermined; we have just three rational numbers
at our disposal to fit 120 monomials.
A solution exists and it is unique: everything is given by
$\zeta(3)\vee\zeta(5)$.

It appears that the QFT expansion ``knows more than we do'' about the
structure of MZVs. Faced with a choice between
$\zeta(5)\vee\zeta(3)$ and $\zeta(3)\vee\zeta(5)$
it rejects the former and takes naive multiplication
as the composition rule for the latter.
For good measure it tells us that the new primitive
is $Z(8)=\zeta(5,3)$, modulo $\pi^8$, with {\em no}
contribution from $\zeta(5)\zeta(3)$. Note that we would
get an unambiguously wrong answer were we to
take the primitive as $\zeta(3,5)$.

The situation becomes yet more interesting at weight 10, where
$\zeta(7)\vee\zeta(3)$ is rejected,
$\zeta(3)\vee\zeta(7)$ is accepted, as a plain multiplication,
and the primitive is
determined to be $Z(10)=\zeta(7,3)+3\zeta^2(5)$, modulo $\pi^{10}$,
with {\em no\/} possibility
of $\zeta(7)\zeta(3)$ appearing in this primitive.
Here one sees the monomials
in the $[\zeta(5)]^2$ term of $S$ generated by the sum
of two terms: $(N(5)-D(5))J_5$ and $K_{10}$, each of which
was predetermined. The former is normalized, as expected, by taking
account of the factor of $1/2$ in pseudo--exponentiation;
for the latter we have only a single rational number,
with which to fit 220 monomials, and the coefficient 3
in $\zeta(7,3)+3\zeta^2(5)$ emerges in every case.

At weight 11, the working out is even more impressive:
$[\zeta(3)]^2\vee\zeta(5)$ and $\zeta(3)\vee\zeta(5,3)$
are present, as naive multiplications;
$(\zeta(5)\zeta(3))\vee\zeta(3)$ is rejected;
and the primitive is given for each of the 286 monomials
as the combination $Z(11)=\zeta(3,5,3)-\zeta(3)\zeta(5,3)$,
in precise accord with analysis of subdivergence--free 7--loop
counterterms~\cite{BK15}.

\section {Experimentum crucis}

Finally, QFT seems to ``know more than we do'' about pushdown of depth--4 MZVs
to depth--2 alternating Euler sums at weight 12.
There still remain several hundred rational numbers to be generated by a
procedure that is now unambiguous: 
$[\zeta(3)]^3\vee\zeta(3)=[\zeta(3)]^3/4$ is combinatorically modified, 
\`a la Kreimer, and $\zeta(7)\vee\zeta(5)=0$,
$\zeta(9)\vee\zeta(3)=0$, are rejected, \`a la Drinfeld-Deligne,
in favour of a pair of primitives.
Taking the first as $Z_1(12)=\zeta(7,5)$ we then determine the second,
$Z_2(12)$, modulo the first, and modulo $\pi^{12}$. 
It is forced to contain all 10 of the remaining MZV structures
in a basis that includes $\zeta(4,4,2,2)$. Moreover the
coefficients of these terms have numerators of typically
10 digits.

When we push $\zeta(4,4,2,2)$ down to depth 2,
via its empirical PSLQ relation to $\zeta(\overline9,\overline3)$,
some sanity emerges, since the terms in $Z_2(12)$ now
involve only double sums and products of two single sums. Yet
still there are apparently grotesque rational coefficients.
Finally, we discovered that this seemingly random assortment of
terms is simply given by~(\ref{other}), with 3 terms whose
coefficients are $\pm1$, though the simplicity of
$\zeta(\overline7)\zeta(\overline5)=945\zeta(7)\zeta(5)/1024$
would have been disguised had we not transformed to a product of
{\em alternating\/} sums.

It is as if QFT were taunting us with our ignorance of
the mapping between diagrams and numbers that results
from the Feynman rules. A vast quantity of data, collected
by painfully inadequate methods, collapses to an amazingly
simple answer. We are, physicists and mathematicians
alike, stumbling on the edge of a structure that is
far more refined than the clumsy methods by which we investigate it.
In this particular instance, it was the sheer volume of apparently
random -- yet in fact gracefully proportioned -- data,
from many numerical instances of self--energy dressings,
that led to an answer to the question posed in the title
of this contribution.

\section{Acknowledgements}

This work began at Boston University and was continued at
the Erwin Schr\"odinger Institute, Wien.
I thank Dirk Kreimer for his unquenchable optimism and
Johannes Blumlein for getting me to recount an arduous journey.
I apologize for not knowing of any more reasonable way of presenting it.

\end{document}